\documentclass{article}

\PassOptionsToPackage{preprint}{neurips_2025}
\PassOptionsToPackage{numbers, compress}{natbib}

\usepackage{neurips_2025}
\usepackage{lipsum}

\usepackage[utf8]{inputenc} 
\usepackage[T1]{fontenc}    
\usepackage{hyperref}       
\usepackage{url}            
\usepackage{booktabs}       
\usepackage{amsfonts}       
\usepackage{nicefrac}       
\usepackage{microtype}      
\usepackage{xcolor} 
\usepackage{graphicx}
\usepackage{amsmath}
\usepackage{multirow}

\newcommand{\Sref}[1]{Section \ref{#1}}

\newcommand{\fref}[1]{Figure~\ref{#1}}

\newcommand{\tref}[1]{Table~\ref{#1}}
\newcommand{\sref}[1]{~\Sref{#1}}

\newcommand{\ours}{TTS-CtrlNet}

\def\maketitlesupplementary
   {
   \newpage
       
        {\centering
        \Large
        \textbf{TTS-CtrlNet: Time varying emotion aligned
text-to-speech generation with ControlNet}\\
        \vspace{0.5em}Supplementary Material \\
        \vspace{1.0em}
   }}

\title{TTS-CtrlNet: Time varying emotion aligned text-to-speech generation with ControlNet}

\usepackage{kotex} 
\begin{document}

\maketitle

\begin{abstract}
  Recent advances in text-to-speech (TTS) have enabled natural speech synthesis, but fine-grained, time-varying emotion control remains challenging. Existing methods often allow only utterance-level control, require full model fine-tuning with a large emotion speech dataset, which can degrade performance. Inspired by \cite{zhang2023addingconditionalcontroltexttoimage}, we propose the first ControlNet-based approach for controlling flow-matching TTS (\ours{}), which freezes the original model and makes a trainable copy of it to process additional condition. We show that \ours{} can boost the pretrained large TTS model by adding intuitive, scalable, and time-varying emotion control while inheriting the ability of the original model (e.g., zero-shot voice cloning \& naturalness). Furthermore, we provide practical recipes for adding emotion control: 1) optimal architecture design choice with block analysis, 2) emotion-specific flow step, and 3) flexible control scale. 
  Experiments show that ours can effectively adds an emotion controller to existing TTS, and achieves state-of-the-art performance with emotion similarity scores: Emo-SIM and Aro-Val SIM. Project page is available at: \href{https://curryjung.github.io/ttsctrlnet_project_page}{https://curryjung.github.io/ttsctrlnet\_project\_page}
\end{abstract}
\section{Introduction}

Recent advances in text-to-speech (TTS) technology have enabled highly natural and expressive speech synthesis, allowing users to generate lifelike utterances from arbitrary text \cite{zhao2023emotion, le2024voicebox, kanda2024making, Eskimez2024E2TE, chen2024f5ttsfairytalerfakesfluent, du2024cosyvoice, wang2017tacotron, ren2019fastspeech}. As these systems mature, user demand for greater controllability, especially over emotional expression, has grown rapidly. While various approaches have been proposed to condition TTS on emotion, most existing methods \citep{zhao2023emotion,tang2024ed,guo2023emodiff,tang2023emomix,zhou2022speech,tang2023qi,lei2022msemotts,shin2022text,lee2017emotional,li2021controllable,cai2021emotion,cho24_interspeech} support only utterance-level control and lack the ability to finely modulate time-varying emotional cues within an utterance. This limitation is particularly critical in applications such as speech-to-speech translation, where accurate emotional transfer aligned over time is essential. In addition, most of them are trained on a limited number of speakers and samples (in an extreme case, only 1 speaker) due to the high cost of collecting a labeled emotion speech dataset, resulting in a lack of control over emotions with arbitrary speakers. This significantly hinders their practical applicability.

\citet{wu2024laugh} shows remarkable success in time-varying emotion conditioning by fine-tuning pre-trained large-scale TTS. However, it 1) requires a huge training cost while fine-tuning the entire pre-trained model with an 87k hours of large-scale speech dataset, which includes 27k hours of in-house emotion dataset, 2) compromises the performance of the pre-trained model, and 3) cannot intuitively adjust emotion intensity.

To address the problem, we first adapt the ControlNet paradigm to flow-matching-based TTS
in order to synthesize time-varying emotions of any zero-shot speaker.
Inspired by \citet{zhang2023addingconditionalcontroltexttoimage}, we freeze the original large-scale pre-trained model and
utilize it as an encoder to leverage its deep prior knowledge in speech synthesis.
The original model and its trainable copy are connected with zero-convolution. It minimizes damage during fine-tuning by progressively increasing the controlling features from a zero tensor. By training the control network while freezing the backbone TTS model, our approach preserves the ability of the original model and enables emotion control with a relatively small amount of unlabeled speech data ($\sim$ 400 hours) compared to the data used for the original model. In addition, we provide practical recipes for effective emotion control: 1) analyzing layer-wise impact on WER, 2) identifying optimal flow step interval for emotion-specific condition, and 3) using the proper temporal window required to capture emotion and 3) flexible interpolation between behaviors of the original model and ControlNet-applied version using a control scale. 

In conclusion, we show that it is possible to efficiently and effectively incorporate desired control capabilities into a large pre-trained system without modifying its original parameters, and without requiring extensive computational resources or costly labeled data.

Our contributions are as follows. 

\begin{itemize}
\item  We first introduce a ControlNet-based conditioning method for large-scale \& flow-matching TTS, enabling fine-grained, time-varying emotion control with zero-shot TTS.

\item  Our approach preserves the quality, zero-shot capability and naturalness of the original TTS model by freezing its parameters during the control network training with careful design choices.

\item We propose effective training and inference strategies to maximize emotional expressiveness with 1) minimal computational cost and 2) a balance between word error and emotion control.

\end{itemize}




\section{Related work}

\subsection{Emotion-controllable text-to-speech}
There have been a number of zero-shot TTS models \cite{chen-etal-2024-f5tts, Eskimez2024E2TE, le2024voicebox} that enable voice cloning using a short reference audio clip. During the voice cloning stage, these models can transfer the emotion expressed in the reference audio, allowing for natural emotion control. \cite{cho24_interspeech} achieves emotion control by incorporating an additional emotion condition into the model.
While previous studies have demonstrated the ability to control emotion at the utterance level, achieving fine-grained, time-varying emotion control remains a significant challenge.
\cite{wu2024laugh} supports time-varying emotional control, it requires full model finetuning, which incurs a substantial computational cost.




\subsection{Controlling large-scale generative model}
Both flow-matching \& diffusion models are a family of score-based generative models which synthesize the target distribution by iteratively denoising random Gaussian noise. Accompanied by diffusion transformer \cite{rombach2021highresolution, baevski2020wav2vec20frameworkselfsupervised, peebles2023scalablediffusionmodelstransformers}, they achieve huge success in scaling the generative model in both image \cite{rombach2021highresolution, Rombach_2022_CVPR, Peebles2022DiT, flux2024, esser2403scaling} and audio modalities, enabling product-level models.
Based on the success, \cite{zhang2023addingconditionalcontroltexttoimage, ye2023ip-adapter} propose to add controllability to the pre-traineded model while leveraging the prior of the model with fine-tuning. \cite{zhang2023addingconditionalcontroltexttoimage} chooses a carefully designed conditional network (ControlNet), by making a trainable copy only at downblocks of the original UNet and connecting it with the original branch through zero-convolution. It leads to huge success in spatial control.
On the other hand, \cite{avrahami2024stable} analyzes the influence of each layer in a DiT trained via flow matching, identifies the vital layers, and demonstrates the possibility of training-free editing during inference.


 \subsection{Flow matching}
 Flow matching \cite{lipman2023flow} learns a vector field $\mathbf{v}_t(\mathbf{x};\theta)$ which transfer the known distribution (e.g., Gaussian noise) to the target distribution $p_1$, approximating the data distribution $q$ with a neural network $\theta{}$. In conditional flow matching (CFM), we have flow map $\boldsymbol{\psi}_t(\mathbf{x}) = \sigma_t(\mathbf{x}_1) x + \mu_t(\mathbf{x}_1)$ conditioned on $\mathbf{x}_1$. With $\mu_0(\mathbf{x}_1) = 0$ and $\sigma_0(\mathbf{x}_1) = 1$, $\mu_1(\mathbf{x}_1) = \mathbf{x}_1$ and $\sigma_1(\mathbf{x}_1) = 0$, a vector field of the flow is $d\boldsymbol{\psi}_t(\mathbf{x}_0)/dt = \mathbf{u}_t\left(\boldsymbol{\psi}_t(\mathbf{x}_0) \mid \mathbf{x}_1\right)$. Then, CFM loss becomes :
 

\begin{equation}
     \mathcal{L}_{\text{CFM}}(\theta) = \mathbb{E}_{t,\, q(\mathbf{x}_1),\, p(\mathbf{x}_0)} \left\| \mathbf{v}_t(\boldsymbol{\psi}_t(\mathbf{x}_0)) - \frac{d}{dt} \boldsymbol{\psi}_t(\mathbf{x}_0) \right\|^2.
\end{equation}

Optimal transport further provides the OT-CFM loss:


\begin{equation}
\mathcal{L}_{\text{CFM}}(\theta) = \mathbb{E}_{t,\, q(\mathbf{x}_1),\, p(\mathbf{x}_0)} \left\| \mathbf{v}_t\left((1 - t)\mathbf{x}_0 + t\,\mathbf{x}_1\right) - (\mathbf{x}_1 - \mathbf{x}_0) \right\|^2,
\end{equation}

where the flow $\boldsymbol{\psi}_t(\mathbf{x}_0)$ is interpolation between $\mathbf{x}_0$ and $\mathbf{x}_1$ with flow step $t \in [0,1]$. With the learned $\mathbf{v}t(\cdot;\theta)$, we synthesize $x_1$ from randomly sampled $\mathbf{x}_0 \sim N(0,I)$ by solving the ordinary differential equation (ODE) with the predicted flow $d\boldsymbol{\psi}_t(\mathbf{x}_0)/dt$. $\mathbf{x}_0$ is progressively transferred into $\mathbf{x}_1$ with the number of function evaluations (NFE) from 0 to 1.

\subsection{Speech emotion recognition}
Speech Emotion Recognition (SER) has received increasing attention in recent years, with diverse modeling approaches proposed to better capture emotional cues from speech \cite{parthasarathy17_interspeech,mirsamadi2017aser,ghosal2019dialoguegcngraphconvolutionalneural}.
One prominent line of work \cite{wagner2023dawn} leverages self-supervised learning to pretrain models on large-scale, unlabeled speech corpora. Models like wav2vec 2.0 \cite{baevski2020wav2vec20frameworkselfsupervised} are trained to learn contextualized audio representations without the need for emotion labels. These pretrained models are then adapted to downstream SER tasks by attaching a regression head or classifier and fine-tuning on emotion-labeled datasets. This strategy is particularly effective when labeled emotion data is scarce.

Since emotion labels are typically provided at the utterance level, the entire speech segment is passed through the pretrained encoder, and the resulting feature sequence is mean-pooled to obtain a single fixed-length representation. A regression head is trained on top of this representation to predict the corresponding emotion label for the utterance.






\begin{figure}[t]
  \centering
\includegraphics[width=0.7\linewidth]{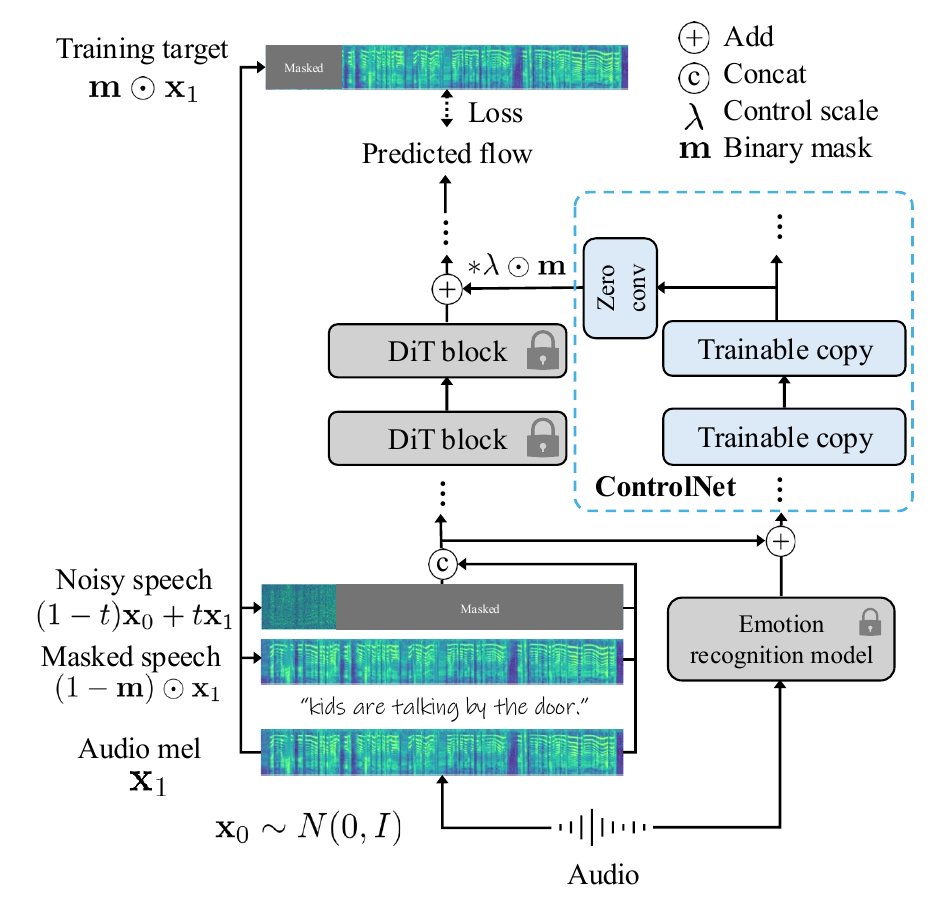}
  \caption{\textbf{Overview of \ours} Controlling signal is processed through ControlNet and fed into the subset of blocks in original model.}
  \label{fig:architecture}
\end{figure}

\section{Method}

\subsection{Pipeline}
In \fref{fig:architecture}, we show an overview of \ours{}. Ours consists of two branch: 1) Original TTS branch and 2) emotion conditioning branch (ControlNet). 

\subsubsection{Architecture}
Inspired by \citet{zhang2023addingconditionalcontroltexttoimage}, \ours{} builds upon a trainable copy of the pre-trained text-to-speech model and connecting the copied and the original model with zero-convolution. 
The zero-convolution is initialized to output a zero tensor, thereby preserving the behavior of the original model while allowing the model to gradually incorporate additional conditioning.
Given a masked audio and a paired transcript, the original text-to-speech model \cite{chen-etal-2024-f5tts} learns an infilling task by predicting the masked part. Specifically, it requires a pair $(\mathbf{x},y)$ where $\mathbf{x} \in \mathbb{R}^{F \times T}$ denotes an melspectrogram of an audio sample $s$ and $y$ is the corresponding transcript. $F$ is dimension of melspectrogram and $T$ is the sequence length. 
Given an input $\mathbf{x}$, we construct a noisy speech sample as $(1 - t)\mathbf{x}_0 + t\mathbf{x}_1$, and a masked speech sample as $(1 - \mathbf{m}) \odot \mathbf{x}_1$, where $t$ denotes the flow step, and $\mathbf{m} \in \{0, 1\}^{F \times T}$ is a random binary temporal mask.
The noisy speech, masked speech, and transcript are concatenated before entering the model. 
Simultaneously, ControlNet also takes the concatenation as input after being added to emotion $\mathbf{e} \in \mathbb{R}^{\ D_\text{emo}\times T}$, which is extracted using the speech emotion recognition model (SER) \cite{wagner2023dawn} to represent arousal, dominance and valance with three channels. It takes an audio sample $s$ as input and outputs the emotion. 
Following \cite{wu2024laugh}, we reject dominance. We match the channel of emotion embedding with the channel of original DiT input embedding ($F$) by passing through a $1\times1$ convolution layer.

The ControlNet consists of multiple DiT blocks. Since the ControlNet is copied from the original network, each block corresponds to its original block. The output from each ControlNet block is added to the output of the corresponding original output block after passing zero-convolution. Let $\mathcal{F}_{k}(\cdot\, ; \, \theta)$ and $\mathcal{F}_{k}(\cdot\, ; \, \phi)$ be a $k$-th original DiT block and the corresponding ControlNet block, respectively. Then, the final output $Z_\text{new}$ of the $k$-th block in the original network becomes 
\begin{equation}
    \mathbf{Z}_\text{new} = \mathcal{F}_{k}(\cdot\, ; \, \theta) + \mathbf{m} \  \odot \mathcal{F}_{k}(\cdot\, ; \, \phi).
\end{equation}

\subsubsection{Training}
For training, we prepare $(x,y)$ training pairs and the corresponding audio $a$, and update only the ControlNet block parameters $\phi$ with the following flow-matching loss, conditioned by emotion $e$:
\begin{equation}
\mathcal{L}_{\text{CFM}}(\phi) = \mathbb{E}_{t, q(\mathbf{x}_1), p(\mathbf{x}_0)} \left\| \mathbf{v}_t\left((1 - t)\mathbf{x}_0 + t \mathbf{x}_1, \mathbf{e}; \theta, \phi \right) - (\mathbf{x}_1 - \mathbf{x}_0) \right\|^2,
\end{equation}

where $\mathbf{v}_t(\cdot{};\theta,\phi)$ is now the combined model with the original model $\theta$ and ControlNet $\phi$. The whole model is trained to predict the flow-derived target distribution $P\left(\mathbf{m} \odot \hat{\mathbf{x}} \,\middle|\, (1 - \mathbf{m}) \odot \hat{\mathbf{x}}, y, \mathbf{e}\right)$.

\paragraph{Emotion-specific flow step} The original model is trained with the random flow step $t \sim [0,1]$. Meanwhile, we opt to use only a subset of the interval while training with additional emotion embedding: $t \sim [0,t_\text{emo}]$, where  $t_{emo} \in {0,1}$ designate the range of employing ControlNet.
It achieves lower word error rate (WER) and higher emotion similarity. We provide the supporting experiments and ablation study in \sref{sec:emotion_flow_step}.

\paragraph{Selective block}
Rather than connecting all DiT blocks, we select only a subset of ControlNet blocks based on their contribution to performance. Specifically, 
to find the most critical blocks,
we conduct a layer-wise ablation study in \sref{sec:selective_block}. Each DiT block is bypassed via its residual connection, and we evaluate the resulting impact on both WER and speaker similarity. This analysis allows us to identify the most influential layers for stable control.

\subsubsection{Inference}
For inference, we prepare reference melspectrogram $x_\text{ref}$ from the audio sample $s_\text{ref}$, the transcript $y_\text{ref}$ of the audio, target text $y_\text{gen}$, initial noise $\mathbf{x}_0 \sim N(0,I)$, and emotion $\mathbf{e}_\text{ref}=\text{SER}(s_\text{ref})$. The original network takes three inputs $(\mathbf{x}_0,\mathbf{x}_\text{ref}, y_\text{text})$ where $y_\text{text} = y_\text{ref} + y_\text{gen}$ and ControlNet takes fours inputs $(\mathbf{x}_0,\mathbf{x}_\text{ref}, y_\text{text}, \mathbf{e})$.
At the output of each DiT block, we get $\mathbf{Z}_\text{new}$ as follows:
\begin{equation}
    \mathbf{Z}_\text{new} = \mathcal{F}_{k}(\cdot\, ; \, \theta) + \lambda \cdot \mathbf{m} \odot  \mathcal{F}_{k}(\cdot\, ; \, \phi),
\end{equation}
where $\lambda$ represents the control scale. The control scale enables balancing between WER and emotion similarity. We provide the experiment according to various control scale in \sref{sec:control_scale}. The original network outputs a predicted flow field, which is used to solve an ODE. By integrating this flow starting from $\mathbf{x}_0$, we can synthesize speech $\hat{\mathbf{x}}_{1}$.

Since ControlNet is trained only within the interval $[0, t_{\text{emo}}]$, we apply it exclusively during this time range for inference. For the remaining steps, we set $\lambda = 0$.

\subsection{Implementation detail}
We use a wav2vec-based emotion recognition model \cite{wagner2023dawn}, which predicts arousal-valence-dominance. Following \cite{wu2024laugh}, we exclude dominance and use only arousal-valence. Emotion recognition model consists of two subsequent network, wav2vec \cite{baevski2020wav2vec20frameworkselfsupervised} and a regression head. First, we get the token length corresponding to the length of the given audio. The token is interpolated with the emotion window size $W_\text{emo}$ and returns $e \in \mathbb{R}^{D_\text{emo} \times T}$. 
After that, $\mathbf{e}$ goes through $1\times1$ convolution layer, resulting in $\mathbf{e} \in \mathbb{R}^{F \times T}$. We use set 1e-5 as a learning rate and train with two A5000 GPUs with 8000 batch frames (the length of melspectrogram) up to 24000 steps for each configuration.

\section{Experiment}

\subsection{Dataset}
\subsubsection{Training}
Publicly available high-quality emotional TTS datasets are extremely scarce. In contrast to large-scale speech corpora, which typically range from 10K to 100K hours \cite{he2024emilia, zen2019libritts}, existing emotional speech datasets are significantly smaller, often limited to just a few hundred hours \cite{lotfian2017building,richter2024ears,busso2008iemocap}. To overcome this limitation, we constructed a training dataset by combining six datasets \cite{busso2008iemocap, richter2024ears, zhou2021seen, lotfian2017building, nguyen23_interspeech}. Since we utilized all datasets with different characteristics, certain criteria were required to curate the training dataset. For the EARS dataset \cite{richter2024ears}, we removed audio files exceeding 3 minutes because the baseline could not handle the audio length. Files consisting solely of non-verbal sounds, such as throat clearing, or lacking corresponding transcriptions, were excluded. Similarly, in the case of the Expresso dataset \cite{nguyen23_interspeech}, multiple speakers' audio found in conversational datasets were omitted. Approximately 400 hours of data remained available for training. All files are upsampled or downsampled to 24kHz to meet the input file specifications of the baseline model.

\subsubsection{Evaluation}
\paragraph{EMO-Change} We follows \cite{wu2024laugh} to implement the EMO-Change dataset using the RAVDESS dataset \cite{livingstone2018ryerson} to evaluate the model’s ability to synthesize speech with nuanced emotional transitions, which features English speech samples labeled with emotions such as calm, happy, sad, angry, fearful, surprised, and disgusted. RAVDESS provides two transcripts -\textit{“kids are talking by the door”} and \textit{“dogs are sitting by the door”}- each spoken with various emotional intensities. For the EMO-Change dataset, we randomly choose two \textit{“kids are talking by the door”} utterances, each expressing a different emotion, remove any silent segments, and concatenate them to form a single audio prompt. During evaluation, this concatenated audio is used as the emotional reference, while the repeated sentence \textit{“dogs are sitting by the door dogs are sitting by the door”} serves as the text prompt for the zero-shot TTS model. This design tests the model’s capability to reproduce the emotional shifts from the audio prompt in its generated speech, while maintaining the target text and speaker identity.
\paragraph{JVNV S2ST} For speech-to-speech translation, we follow the Japanese-to-English speech-to-speech translation (S2ST) experimental protocol of \cite{wu2024laugh} to evaluate the emotion transferability of zero-shot TTS models in a cross-lingual setting.
Specifically, we use the JVNV dataset, which contains emotionally expressive Japanese speech from four speakers (two male and two female) covering six emotions: anger, disgust, fear, happiness, sadness, and surprise, as well as diverse nonverbal vocalizations. For S2ST translation, we first perform speech-to-text translation on the Japanese utterances to obtain English transcripts. These transcripts served as text prompts, while the original Japanese audio was used as the audio prompt for the zero-shot TTS model, from which we extracted both nonverbal and emotion embeddings. Since our base model \cite{chen-etal-2024-f5tts} is not trained on Japanese and English within a model, we transform the Japanese transcript into its phonetic representation using  Romaji\footnote{https://en.wikipedia.org/wiki/Romanization\_of\_Japanese}. Given a transcript and the corresponding source audio, the model generates English speech that preserves the speaker’s identity and emotional characteristics from the source. We evaluate the similarity of speaker and emotion between the original Japanese speech and the synthesized English output using objective and subjective metrics, as described in the following section.

\subsection{Evaluation metric}
\paragraph{Objective metric} Following the evaluation protocol of \cite{wu2024laugh}, we utilize the same set of metrics to assess both the intelligibility and emotional controllability of the generated speech. We applied Whisper-Large \cite{10.5555/3618408.3619590} for automatic transcription and computing the corresponding ground truth texts. We express word error rate (WER) in percentage. To assess prosodic similarity, we utilized AutoPCP \cite{barrault2023seamless}, a sentence-level metric that estimates the prosodic alignment between speech signals. We computed the AutoPCP score between the generated audio and the reference audio using AutoPCP multilingual v2.\footnote{\href{https://github.com/facebookresearch/seamless_communication}{https://github.com/facebookresearch/seamless\_communication}} We adopt Emo-SIM and Aro-Val SIM to evaluate the emotional content of the reference audio is well reflected in the synthesized speech. Emo-SIM measures similarity based on discrete emotional states using frame-level embeddings extracted by the Emotion2Vec model \cite{ma-etal-2024-emotion2vec}. Aro-Val SIM evaluates alignment in continuous emotion space—arousal and valence—estimated using a sliding window approach following \cite{wagner2023dawn}. Together, these metrics provide a comprehensive assessment of emotional expressiveness in the synthesized speech.


\paragraph{Subjective metric} We employed the following subjective evaluation metrics to assess the perceptual quality of the synthesized speech. SMOS (Speaker Similarity Mean Opinion Score) measures the perceived similarity between the speaker prompt and the generated speech on a 5-point Likert scale, ranging from 1 (not at all similar) to 5 (extremely similar). NMOS (Naturalness Mean Opinion Score) evaluates the naturalness of the synthesized speech, where 1 indicates “bad” and 5 indicates “excellent.” EMOS (Emotion Mean Opinion Score) assesses the emotional similarity between the reference audio prompt and the generated speech, rated from 1 (not at all similar) to 5 (extremely similar).

\begin{figure}[t]
  \centering
\includegraphics[width=0.9\textwidth]{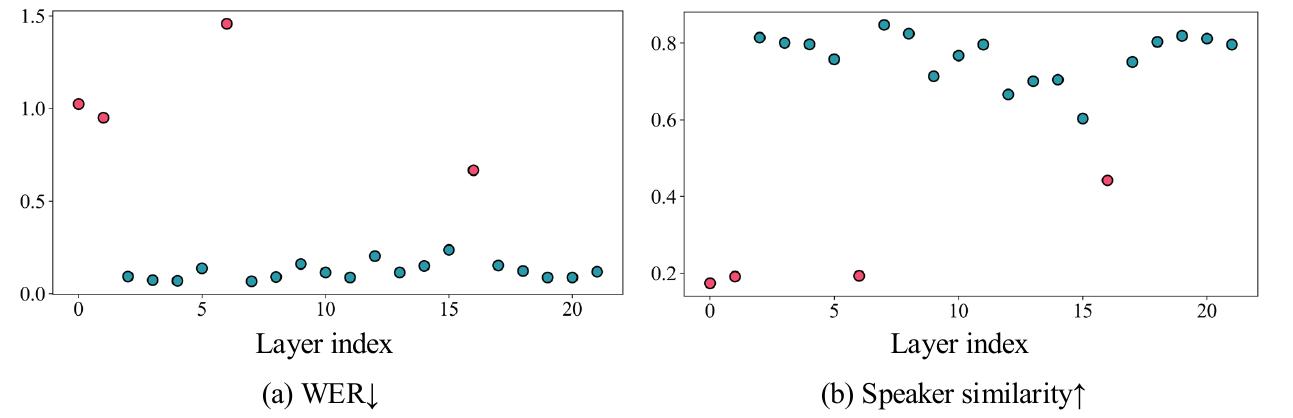}
  \caption{\textbf{Layer ablation with quantitative results} We measure WER and speaker similarity after skipping each block during inference and find that some blocks greatly increase WER and decrease speaker similarity when skipped.}
  \label{fig:layer_ablation}
\end{figure}

\subsection{Ablation study}
For the ablation study, we used the RAVDESS dataset as the evaluation data \cite{livingstone2018ryerson}.

\subsubsection{Selective block}
\label{sec:selective_block}
In this section, we propose a block-level analysis with transformer-based TTS models. 
Leveraging the structural property of DiT, Stable Flow \cite{avrahami2024stable} analyzes the contribution of each block in a flow-matching-based text-to-image model, ultimately enhancing controllability through a training-free approach. In a similar vein, but with a new perspective, we investigate the transformer blocks in F5-TTS, which consist of 22 sequentially connected residual blocks.
In \fref{fig:layer_ablation} (a), we evaluate the impact of each block by skipping it during inference and measuring the resulting word error rate (WER). We observe that skipping certain blocks significantly increases WER. Moreover, as shown in \fref{fig:layer_ablation} (b), the same blocks whose removal increases WER also lead to a notable drop in speaker similarity. This indicates that these blocks are critical for preserving both speaker identity and textual fidelity.

Based on this insight, we exclude such crucial blocks from being connected to ControlNet. As shown in \tref{tab:selective_blocks}, our strategy yields lower WER while achieving high speaker \& emotion similarities, demonstrating the effectiveness of block-wise selection.

\begin{figure}[t]
  \centering
\includegraphics[width=0.7\textwidth]{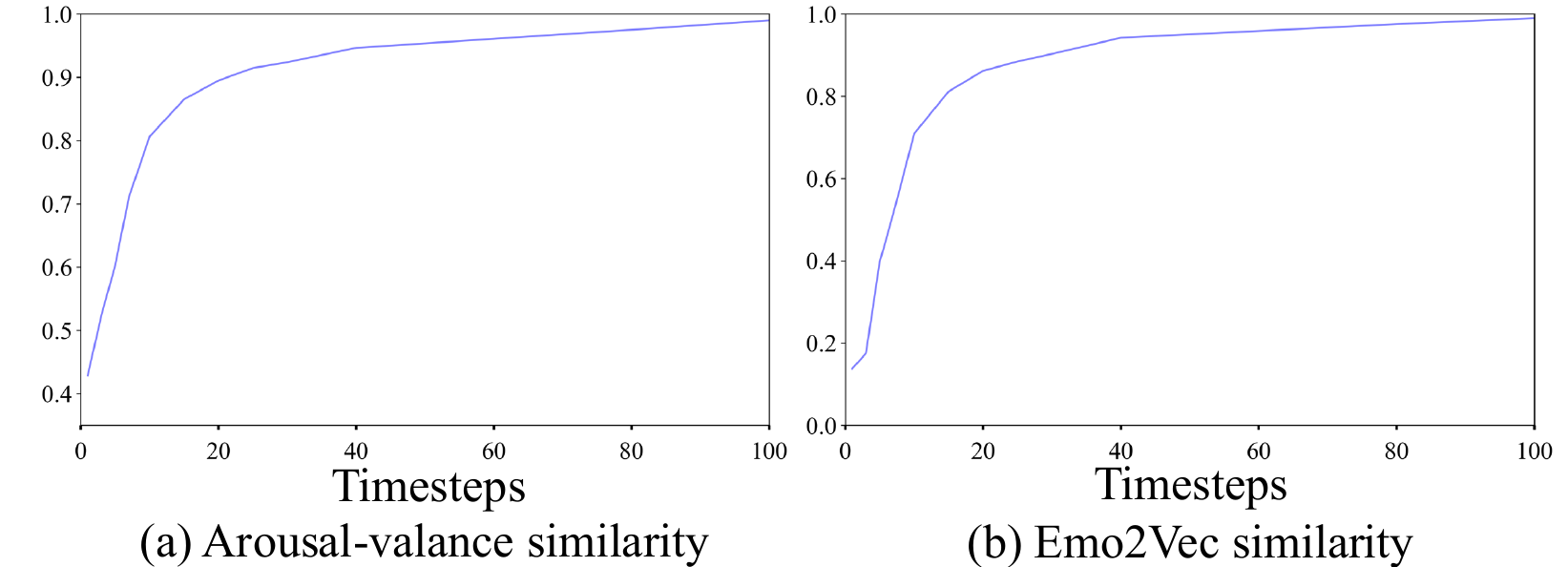}
  \caption{\textbf{Flow step experiments}}
  \label{fig:emo_sim_recon_ablation}
\end{figure}
\begin{table}[]
\caption{\textbf{Training flow step interval ablation study}}
\label{tab:train_timestep}
\centering

\begin{tabular}{ccccc}
\hline
timesteps    & WER (\%) $\downarrow$ & Emo-SIM $\uparrow$ & Aro-Val SIM $\uparrow$ \\ \hline
{[}0,1{]}     & 0                   & 0.389                    & 0.674                      \\ 
{[}0,0.1{]}   & 1.9                   & 0.565                    & 0.876                      \\ \hline

\end{tabular}
\end{table}

\begin{table}[t]
\caption{\textbf{Emotion window size ablation study}}
\label{tab:train_emo_interval}
\centering
\resizebox{0.7\linewidth}{!}{\begin{tabular}{ccccc}
\hline
Emotion window size & WER (\%) $\downarrow$ & Emo-SIM $\uparrow$ & Aro-Val SIM $\uparrow$ \\ \hline
1                   & 4.7                   & 0.500                    & 0.836                               \\
30                   & 1.9                   & 0.565                    & 0.876            
                 \\ \hline
\end{tabular}
}
\end{table}

\subsubsection{Emotion-specific flow step}
\label{sec:emotion_flow_step}
We identify the emotion-specific flow step interval, which is the range of flow steps where the emotion of the generated speech is determined. In a flow-matching-based model, each flow step incrementally evolves a sample from a known prior distribution (Gaussian noise) toward the target distribution (e.g., natural speech). At early flow steps, the sample remains close to Gaussian noise, so even minor perturbations can significantly affect the output. Conversely, as the flow step approaches 1-meaning the sample is near the target distribution, perturbations have minimal impact on the final output.

In this section, we present experiments designed to pinpoint the flow step interval where emotion is established. We prepare multiple \textit{(audio sample, transcript)} pairs and, following the conditional flow matching training objective, interpolate between random Gaussian noise and audio at various flow steps to obtain intermediate flow maps. For each intermediate flow map, we resample and measure the emotional difference from the original audio. Emotion is quantified using arousal-valence (aro-val) and emotion similarity (emo-sim) metrics.

As illustrated in \fref{fig:emo_sim_recon_ablation}, when the flow step $t$ is close to 1, there is little change in emotion. However, if we select a flow step $t$ near the Gaussian noise, the model fails to reconstruct the original audio, producing outputs with low emotion similarity. Through this process, we identify the emotion-specific flow step interval-where emotion similarity changes rapidly, $t \in \{0,t_\text{emo}\}$. Utilizing this interval during training and inference allows us to reduce word error rate (WER) and improve efficiency by excluding regions where emotion does not change from training.  In \tref{tab:train_timestep}, using the whole time step struggles to transfer emotion. 

\subsubsection{Emotion window size}
\label{sec:emotino_window_size}
Leveraging Wav2vec \cite{baevski2020wav2vec20frameworkselfsupervised} pretrained with a large corpus of unlabelled audio dataset by training only a regression head presents a promising approach for emotion recognition, especially in scenarios where labeled emotion data is scarce. Wav2vec produces a sequence of output tokens whose length is proportional to the input audio sequence. By averaging these token representations and passing the result through a regression head, we can predict the emotion of an entire audio clip.
A naive approach to obtain time-varying emotion is to feed each Wav2vec output token directly into the regression head, thereby producing an emotion label for each token. However, this method leads to a loss of emotional information and results in increased word error rate (WER). To address this, we adopt a window sliding interpolation strategy on the Wav2vec features. By applying this technique, we generate smoother and more context-aware features, which are then used as input to ControlNet. This approach preserves emotional information and decrease WER in \tref{tab:train_emo_interval}.

\subsubsection{Control scale}
\label{sec:control_scale}
During training, the control scale is fixed at 1.0, but during inference, it can be freely adjusted. This flexibility allows the model to interpolate between the behaviors of the original model and the ControlNet-enhanced model.
In this experiment, we investigate the impact of the control scale hyperparameter $\lambda$ on the performance of our TTS system. To ensure that the observed effects were solely attributable to the control scale, we apply different values of $\lambda$ to a vanilla F5-TTS model without performing any additional fine-tuning. Specifically, setting $\lambda = 0$ activates only the base F5-TTS model, while $\lambda = 1.0$ fully integrates the ControlNet \cite{zhang2023addingconditionalcontroltexttoimage} output. This design enables a controlled analysis of how varying degrees of ControlNet influence affect the model’s performance while preserving the original model's capabilities. We employ several metrics—including WER, AutoPCP, SIM-O, Emo-SIM, and Arousal-Valence SIM—to evaluate the overall performance of the TTS model, with a particular focus on its ability to convey emotions and synthesize expressive speaking styles. 
As shown in the table \ref{tab:ctrl_scale_ablation}, we observe that increasing the control scale $\lambda$ leads to consistent improvements in the AutoPCP, Emo-SIM, and Arousal-Valence SIM metrics. Since these metrics quantify the similarity and consistency of emotional expression between the reference and synthesized speech, this trend suggests that the control scale effectively enhances the emotional expressiveness of the generated audio. However, we also observe a substantial increase in WER as $\lambda$ increases. We attribute this to a trade-off between emotional expressiveness and linguistic clarity—stronger emotional modulation may introduce acoustic variations that degrade phoneme-level precision, thereby impacting word recognition accuracy.

\begin{table}[t]
\caption{\textbf{Selective blocks ablation study}}
\label{tab:selective_blocks}
\centering
\resizebox{0.7\linewidth}{!}{\begin{tabular}{ccccc}
\hline
Blocks               & SIM-o $\uparrow$ & WER (\%) $\downarrow$ & Emo-SIM $\uparrow$ & Aro-Val SIM $\uparrow$ \\ \hline
Full                 & 0.630                & 8.9                   & 0.450                    & 0.746                     \\
Selective blocks & 0.684                & 0                   & 0.565                    & 0.872                      \\ \hline
\end{tabular}
}
\end{table}

\begin{table}[]
\caption{\textbf{Results with control scale}}
\label{tab:ctrl_scale_ablation}
\centering
\resizebox{0.85\linewidth}{!}{\begin{tabular}{cccccc}
\hline
Control scale & SIM-o $\uparrow$ & WER (\%) $\downarrow$ & Emo-SIM $\uparrow$ & Aro-Val SIM $\uparrow$ & AutoPCP $\uparrow$ \\ \hline
0.0           & 0.579                & 2.9          & 0.692            & 0.845   & 3.62    \\
0.1           & 0.589                &0.63   & 0.724       & 0.864        & 3.68   \\
0.3           & 0.590                & 6.56        & 0.735          & 0.887       & 3.69  \\
0.5           & 0.595                & 7.92         & 0.734     & 0.892     & 3.69      \\
1.0           & 0.593                & 9.58           & 0.742        & 0.892    & 3.67                  \\ \hline
\end{tabular}}
\end{table}

\subsection{Comparisons}
We compare \ours{} with flow-matching based zero-shot TTS models: VoiceBox \cite{le2024voicebox}, ELaTE \cite{kanda2024making}, EmoCtrl-TTS \cite{wu2024laugh}, F5-TTS \cite{chen2024f5ttsfairytalerfakesfluent}. We also include SeamlessExpressive \cite{barrault2023seamless}. Since the most of the model is not publicly available, we use the values reported by Emoctrl-TTS \cite{wu2024laugh} except for F5-TTS. For a fair comparison, we faithfully follow the evaluation protocol to implement JVNV S2ST and EMO-Change dataset. 
\paragraph{Objective evaluation}
In \tref{tab:jvnv_emochange_quan_vs_competitor}, we show the quantitative comparison with JVNV S2ST and EMO-Change datset
 In both JVNV S2ST \& Emo-Change dataset, we achieve the best emotion similarity, including Emo-SIM and Aro-Val SIM, which supports that \ours{} effectively conveys the emotion in a reference audio. We also achieve competitive text fidelity performance in EMO-Change dataset. Since F5-TTS is only trained with english and chinese in Emilia101K \cite{he2024emilia}, it is supposed to take input from the same languages shown during training, but not from japanese. The discrepancy may cause a lower WER compared to the reported value in the official repo of F5-TTS. Please note that we build \ours{} up on F5-TTS. We can find that \ours{} do not harm the text fidelity of the original model in EMO-change, which is enligsh dataset. We also preserve the speaker similarity of our base model in both datasets. Despite being trained on a relatively small dataset (400 hours), our model achieves strong performance across multiple metrics, demonstrating the efficiency and effectiveness of our approach.

\paragraph{Subjective evaluation}
We conducted a comparative evaluation of F5-TTS and \ours{} across SMOS, NMOS, and EMOS with the JVNV S2ST and EMO-Change datasets. For each dataset, we collect 20 audio samples per model. Each MOS category is rated by a panel of 10 participants.

The MOS results are well aligned with our quantitative results in \tref{tab:jvnv_emochange_quan_vs_competitor}. Specifically, both SMOS and NMOS indicate that \ours{} effectively preserves the speaker similarity and naturalness of the base model. Furthermore, the observed improvement in EMOS demonstrates that our approach enables seamless and robust emotion control. We exclude other competitors due to the lack of a sufficient number of speech samples for comparison.


\begin{table}[]
\centering
\caption{\textbf{Quantitative comparison with JVNV S2ST and EMO-change dataset.}}
\label{tab:jvnv_emochange_quan_vs_competitor}
\resizebox{\linewidth}{!}{%
\begin{tabular}{lllccccc}
\hline
Model                        & Init & \multicolumn{1}{c}{Training data (hours)} & SIM-o $\uparrow$ & WER (\%) $\downarrow$ & \multicolumn{1}{l}{AutoPCP $\uparrow$} & Emo-SIM $\uparrow$ & Aro-Val SIM $\uparrow$ \\ \hline
\multicolumn{8}{c}{JVNV S2ST}                                                                                                                                \\ \hline
(B1) SeamlessExpressive      & -    & -                                         & 0.268            & \textbf{1.2}                 & 2.91               & 0.653                & 0.494                  \\
(B2) VoiceBox (reproduction) & -    & LL (60k)                                  & 0.347            & 2.1                 & 2.96               & 0.655                & 0.443                  \\
(B3) ELaTE                   & B2   & LL (60k) + LAUGH (460)                    & 0.441            & 3.8                 & 3.36               & 0.671                & 0.548                  \\
(B4) VoiceBox (fine-tuned)   & B2   & LL (60k) + IH-EMO (27k) + LAUGH (460)     & 0.455            & 3.0                 & 3.17               & 0.659              & 0.470                  \\
(B5) EmoCtrl-TTS             & B2   & LL (60k) + IH-EMO (27k) + LAUGH (460)     & 0.448            & 4.4                 & 3.38                                   & 0.693                & 0.647                  \\
(B6) EmoCtrl-TTS(+)          & B2   & LL (60k) + IH-EMO (27k) + LAUGH (460)     & \textbf{0.497}            & 3.2                 & \textbf{3.50}                                   & 0.697                & 0.643                  \\
(B7) F5-TTS                  & -    & Emilia (95k)                              & 0.459                & 4.5                   & 2.87                                      & 0.684                    & 0.627                      \\
TTS-CtrlNet (Ours)           & B7   & Public emotion speech (400)               & 0.464                & 5.4                   & 2.36                                      & \textbf{0.751}                    & \textbf{0.742}                      \\ \hline
\multicolumn{8}{c}{EMO-change}                                                                                                                                                                                    \\ \hline
(B2) VoiceBox (reproduction) & -    & LL (60k)                                  & 0.600            & 1.2                 & 3.31                                   & 0.685                & 0.663                  \\
(B3) ELaTE                   & B2   & LL (60k) + LAUGH (460)                    & 0.643            & 0.2                 & 3.52                                   & 0.700                & 0.761                  \\
(B4) VoiceBox (fine-tuned)   & B2   & LL (60k) + IH-EMO (27k) + LAUGH (460)     & 0.622            & 1.1                 & 3.31                                   & 0.678                & 0.655                  \\
(B5) EmoCtrl-TTS             & B2   & LL (60k) + IH-EMO (27k) + LAUGH (460)     & 0.671            & \textbf{0.0}                 & 3.45                                   & 0.685                & 0.822                  \\
(B6) EmoCtrl-TTS(+)          & B2   & LL (60k) + IH-EMO (27k) + LAUGH (460)     & \textbf{0.684}            & 0.9                 & 3.44                                   & 0.679                & 0.811                  \\
(B7) F5-TTS                  & -    & Emilia (95k)                              & 0.579            & 2.9                 & 3.62                                   & 0.692                & 0.845                      \\
TTS-CtrlNet (Ours)           & B7   & Public emotion speech (400)               & 0.589                & 0.6                   &\textbf{3.68}                                      &\textbf{0.724}        &\textbf{0.864}                      \\ \hline
\end{tabular}
}
\end{table}



\begin{table}[]
\centering
\caption{\textbf{MOS Comparison under JVNV S2ST and Emo-Change Conditions}}
\label{tab:mos_dual_condition}
\resizebox{0.6\linewidth}{!}{%
\begin{tabular}{lcccccc}
\toprule
\multirow{2}{*}{\textbf{Model}} 
& \multicolumn{3}{c}{\textbf{JVNV S2ST}} 
& \multicolumn{3}{c}{\textbf{Emo-Change}} \\

& \raisebox{-0.5ex}{SMOS} & \raisebox{-0.5ex}{NMOS} & \raisebox{-0.5ex}{EMOS}
& \raisebox{-0.5ex}{SMOS} & \raisebox{-0.5ex}{NMOS} & \raisebox{-0.5ex}{EMOS} \\

\cmidrule(lr){2-4} \cmidrule(lr){5-7}

F5-TTS                & 2.0  & 2.3  & 2.6  & 3.7  & 3.8  & 2.0 \\
\ours & \textbf{2.7}  & \textbf{3.2}  & \textbf{3.4}  & \textbf{3.6}  & \textbf{3.8}  & \textbf{3.3} \\ 

\bottomrule
\end{tabular}%
}
\end{table}


\section{Conclusion}
In this paper, we propose a ControlNet-based method for adding time-varying emotion control to pretrained TTS models. Our approach enables the addition of desired conditions with relatively small public data and low training cost, without the need for full fine-tuning of the base model. We identify an emotion-specific interval within the multiple flow steps of a flow-matching-based model and leverage it during both training and inference to enhance performance. Furthermore, by analyzing the impact of each block within the DiT architecture, we selectively connect ControlNet and the original model, achieving superior results. We also propose an appropriate window size for extracting time-varying emotions from the emotion recognition model, which contributes to more natural and expressive speech synthesis. As a result, our method preserves the naturalness and zero-shot TTS capabilities of the original large-scale model, while significantly improving emotion control. Lastly, our work suggests the possibility of alleviating the burden of increasingly large AI systems by efficiently introducing fine-grained controllability, without relying on extensive computation or costly data.

\paragraph{Limitations} Despite the effectiveness of our proposed approach, it is inherently limited by the capabilities of the underlying speech emotion recognition model (SER). Specifically, our SER cannot accurately recognize non-verbal cues (e.g., laughing, crying), as these are not reliably captured by current emotion recognition methods. For future work, we suggest exploring the integration of alternative condition encoders to support multiple and non-verbal conditions, thereby expanding the expressive range of controllable TTS systems.

\paragraph{Societal Impact}
Emotion-controllable TTS systems have the potential to greatly enhance human-computer interaction by enabling more expressive and empathetic voice-based applications, such as assistive technologies, virtual agents, and language learning tools. However, this technology also raises ethical concerns. Emotion manipulation through synthetic speech could be misused for deceptive purposes, including impersonation, emotional manipulation, or misinformation. As the technology advances, it is important to establish safeguards, such as watermarking, consent mechanisms, and responsible usage guidelines, to ensure it is deployed transparently and ethically that prioritizing user trust and well-being.

{\small
    \bibliographystyle{unsrtnat}
    \bibliography{main}
}

\appendix

\maketitlesupplementary
\setcounter{page}{1}

\section*{Table of Contents}
\begin{itemize}
  \item \nameref{asec:qualitative_results}
  \item \nameref{asec:user_study}
  \item \nameref{asec:emotion_flow_step}
  \item \nameref{asec:emotion_window_size}
\end{itemize}

\section{Qualitative results}
\label{asec:qualitative_results}
We provide qualitative results in \textit{“supple\_demo/index.html”}, which can be accessed after extracting the attached \textit{“supplementary.zip”} file.
Since ELaTe \cite{chen2024f5ttsfairytalerfakesfluent}, Voicebox \cite{le2024voicebox}, and EmoCtrl-TTS \cite{wu2024laugh} are not publicly available,  all the qualitative results of competitors are from the official project page of EmoCtrl-TTS, excluding F5-TTS.

With EMO-Change, the reference audio is a concatenation of two utterances with different emotions, so we expect the generated audio to reflect both emotions in sequence. However, Seamless, ELaTE, Voicebox, and F5-TTS tend to generate speech with a single dominant emotion rather than a sequential transition. In contrast, our method captures both emotions well, aligning them sequentially with natural prosody and speaker similarity comparable to or better than the backbone. EmoCtrl-TTS also performs similarly well. We note that F5-TTS suffers from artifacts while synthesizing high-pitch speeches (e.g., an overly excited female voice). Since \ours uses F5-TTs as a backbone, both F5-TTS and ours suffer from the results of "Happy->Disgusted" (a) with the EMO-Change dataset. 

With JVNV-S2ST, our method effectively reflects time-varying emotional changes. For example, in “angry” (b), the emotion intensifies toward the end of the utterance. The generated speech remains natural, with no degradation in speaker similarity, and often even yields better results. Some outputs show atypical English pronunciation, which is expected since the baseline model, F5-TTS, is trained only on Chinese and English, not Japanese. As a result, when the input audio is in Japanese, traces of Japanese prosody can appear. Our model occasionally shows similar tendencies.

Notably, EmoCtrl-TTS utilizes both the emotion encoder and NV encoder, while ours only uses the emotion encoder. Therefore, please do not account for the transfer of non-verbal cues such as laughter or sobbing.

\begin{figure}[h]
  \centering
\includegraphics[width=0.65\linewidth]{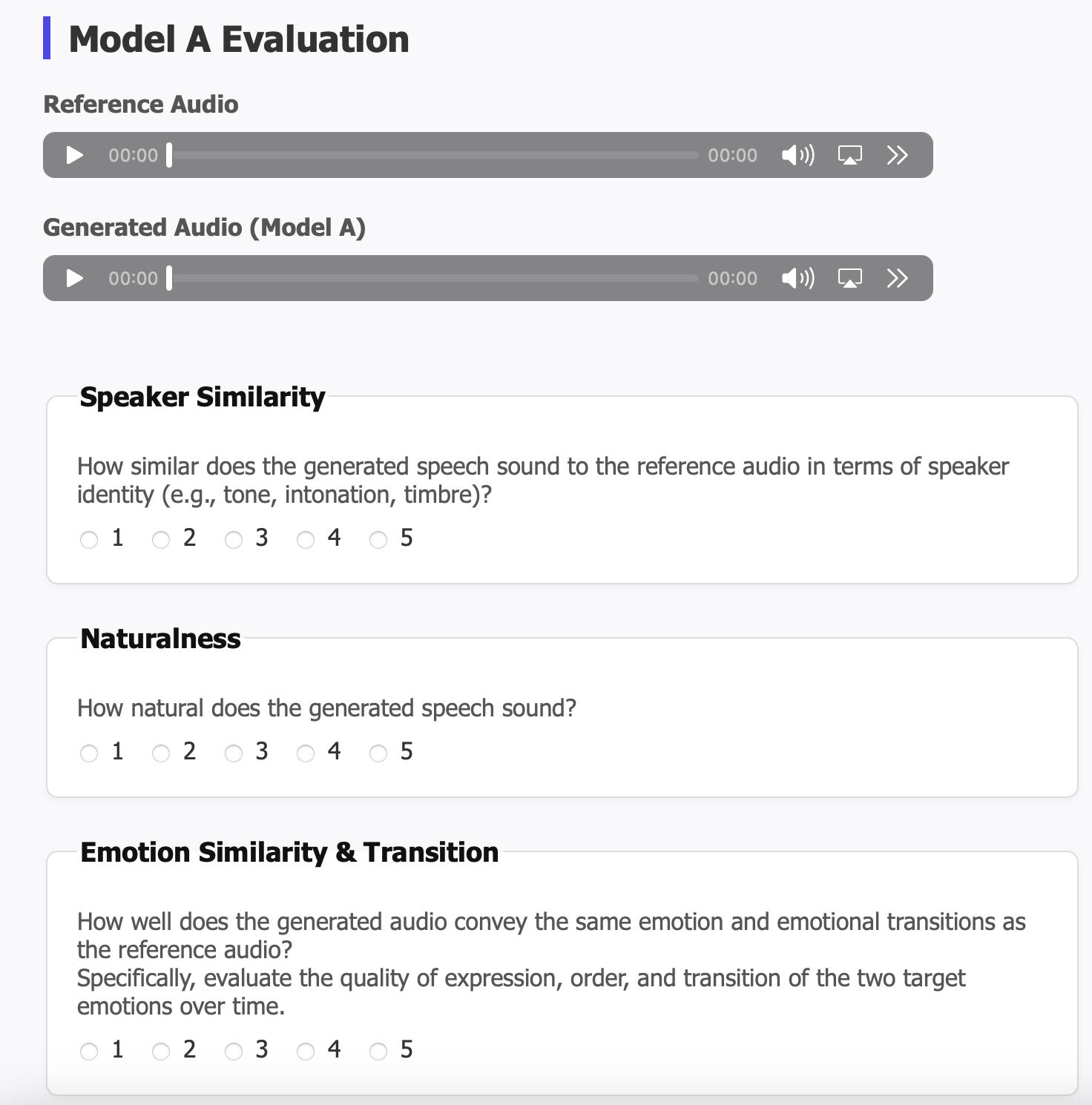}
  \caption{\textbf{Survey page}  }
  \label{fig:survey_page}
\end{figure}

\section{Additional user study}
\label{asec:user_study}
We provide the actual survey page used in our study in \fref{fig:survey_page}. 

In the main paper, we only included a comparison with F5-TTS and excluded other competitors due to the limited availability of their models and speech samples. Since most competitors are not publicly accessible, we could not collect a sufficient number of audio samples for a comprehensive comparison. To address this, we present a qualitative evaluation in this section based on a limited set of samples, using three perceptual metrics: SMOS (Speaker similarity Mean Opinion Score), NMOS (Naturalness Mean Opinion Score), and EMOS (Emotion Mean Opinion Score).

The evaluation is conducted in two independent parts, corresponding to the two datasets described earlier. Part 1 compares 11 Japanese reference audio samples with English audio generated by our model and five baseline models. Part 2 compares English reference samples (8 utterances) against outputs from our model and four baselines. All ratings were collected on a 5-point Likert scale (1–5), and consistent with the main paper, each MOS category was evaluated by 10 participants.

\paragraph{Subjective evaluation} We provide the quantitative results within 6 models in \tref{atab:jvnv_emochange_quan_vs_competitor}. The values in parentheses next to each score represent the 95\% confidence intervals. In terms of EMOS, ours ranks 1st on JVNV-S2ST and 2nd on EMO-Change, showing comparable performance to EmoCtrl-TTS. Considering the confidence intervals, the difference between 1st and 2nd place is not statistically significant. Compared to its baseline model F5-TTS, our method achieves substantial improvements in emotional similarity, with EMOS increasing from 2.91 to 3.55 on JVNV-S2ST and from 2.69 to 4.02 on EMO-Change. Furthermore, ours maintains or improves upon F5-TTS in both SMOS and NMOS, and also outperforms or matches other baselines within the margin of error, indicating strong overall performance across all metrics.

\begin{table}[h]
\caption{\textbf{More quantitative comparison with JVNV S2ST and EMO-change dataset.}}
\label{atab:jvnv_emochange_quan_vs_competitor}
\resizebox{\linewidth}{!}{%
\begin{tabular}{cccccccccc}
\hline
\multirow{2}{*}{ID} & \multicolumn{4}{c}{JVNV S2ST}              & \multicolumn{5}{c}{EMO-Change}               \\ \cline{2-10} 
                    & Model                & SMOS$\uparrow$      & NMOS$\uparrow$      & EMOS $\uparrow$      & Model                & SMOS $\uparrow$    & NMOS $\uparrow$     & EMOS $\uparrow$     &  \\ \hline
B1                  & SeamlessExpressive   & 2.19$_{\pm 0.40}$    & 3.18$_{\pm 0.40}$    & 3.05$_{\pm 0.37}$     &                      &                    &                     &                     &  \\
B2                  & VoiceBox             & 2.81$_{\pm 0.36}$    & 3.06$_{\pm 0.46}$    & 3.35$_{\pm 0.35}$     & VoiceBox             & 4.05$_{\pm 0.25}$  & 3.96$_{\pm 0.27}$    & 2.57$_{\pm 0.43}$    &  \\
B3                  & Elate                & 3.27$_{\pm 0.30}$    & 2.89$_{\pm 0.44}$    & 3.51$_{\pm 0.32}$     & Elate                & 4.26$_{\pm 0.28}$   & 4.26$_{\pm 0.23}$    & 3.52$_{\pm 0.27}$    &  \\
B4                  & EmoCtrl-TTS          & 2.87$_{\pm 0.34}$    & 3.03$_{\pm 0.48}$    & 3.26$_{\pm 0.39}$     & EmoCtrl-TTS          & 4.24$_{\pm 0.24}$   & 4.14$_{\pm 0.22}$    & 4.18$_{\pm 0.21}$    &  \\
B5                  & F5-TTS               & 2.26$_{\pm 0.38}$    & 2.91 $_{\pm 0.47}$   & 2.91 $_{\pm 0.40}$    & F5-TTS               & 3.93$_{\pm 0.25}$   & 3.93$_{\pm 0.25}$    & 2.69$_{\pm 0.40}$    &  \\
--                  & \ours (Ours)         & 2.90$_{\pm 0.36}$    & 3.68$_{\pm 0.34}$    & 3.55$_{\pm 0.40}$     & \ours (Ours)         & 3.96$_{\pm 0.23}$   & 4.10$_{\pm 0.25}$    & 4.02$_{\pm 0.29}$    &  \\ \hline
\end{tabular}
}
\end{table}


\section{More explanation of emotion-specific flow step}
\label{asec:emotion_flow_step}
\sref{sec:emotion_flow_step} finds the flow steps controlled to reflect the reference emotion. We name these steps ``the emotion-specific flow step".
Specifically, we investigate \textit{reconstruction fidelity} where a real audio sample is given. First, we apply interpolation with random noise to the original audio $\mathbf{x}_0$ to compute intermediate latent flows at various flow steps. We then evaluate how well the emotional content can be reconstructed from these flows using two metrics: Arousal-Valence similarity (Aro-Val sim) and Emo2Vec similarity.We define flow steps that result in low emotion similarity as emotion-specific flow steps, since they show poor emotion reconstruction. Our key finding is that emotion control is most effective when applied at these step.
\begin{figure}[h]
  \centering
\includegraphics[width=1.0\linewidth]{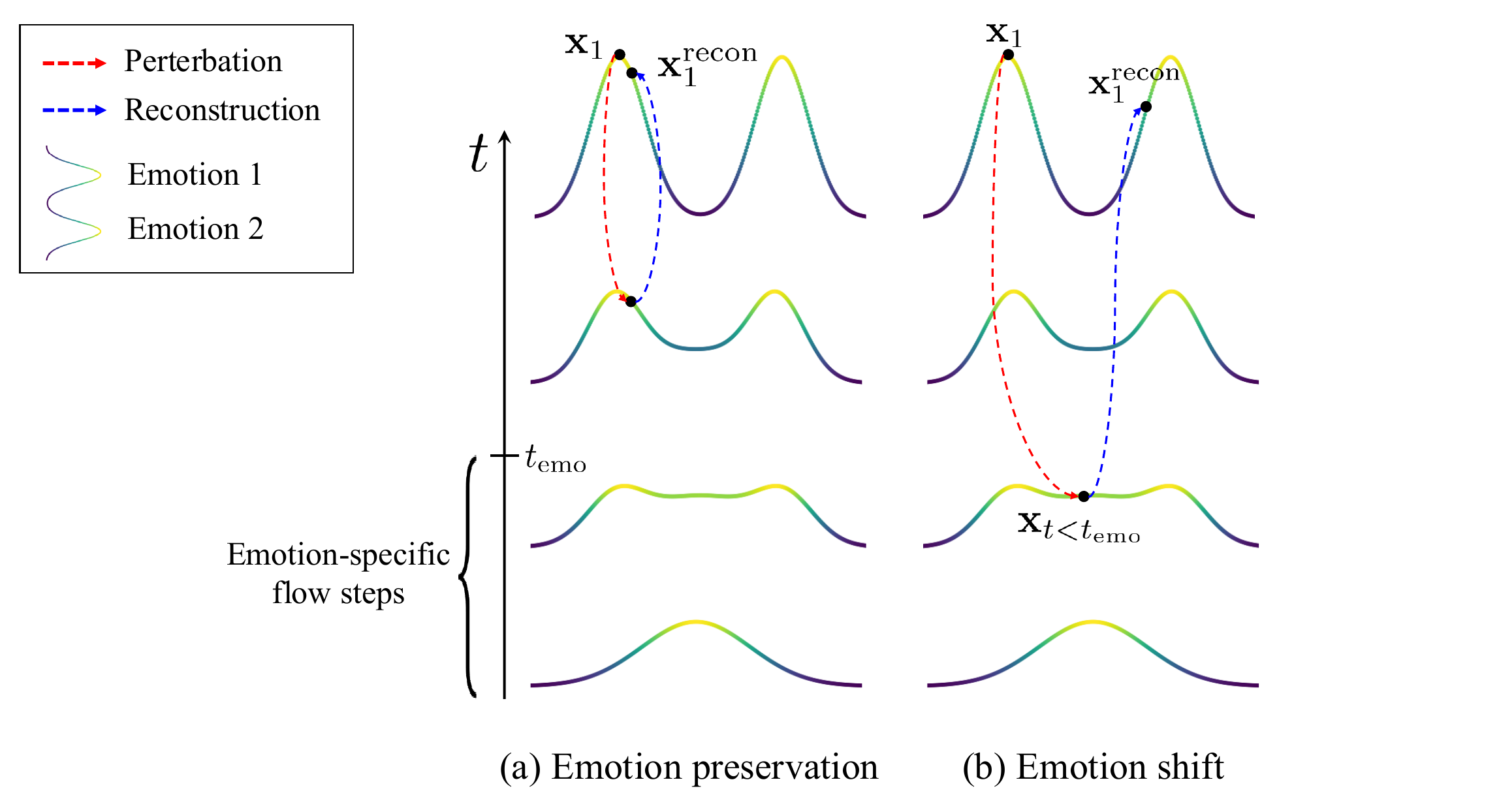}
  \caption{\textbf{Illustration of emotion-specific flow steps.} (a) Slight perturbation of $\mathbf{x}_0$ preserves emotion similarity. 
(b) Perturbation beyond the emotion-specific flow step leads to failed emotion reconstruction and emotion shift.
}
  \label{fig:emotion_flow_step_description}
\end{figure}

\fref{fig:emotion_flow_step_description} (a) shows a scenario where emotion reconstruction succeeds. When perturbing $\mathbf{x}_0$ along a flow path closer to the target emotion distribution (as opposed to pure noise), the reconstructed audio $\mathbf{x}{\text{recon}}$ remains emotionally aligned with $\mathbf{x}_0$, resulting in high similarity.
\fref{fig:emotion_flow_step_description} (b) shows a failure case. Perturbing $\mathbf{x}_0$ in a noise-like direction moves the intermediate representation away from the correct emotion, leading $\mathbf{x}_{\text{recon}}$ to align with a different emotional distribution. 

\paragraph{Evaluation samples} To evaluate this, we sample 120 speech audio clips from the RAVDESS dataset, ensuring a diverse range of emotional expressions. We use F5-TTS as the TTS model and, for each audio sample, performed interpolation with six different random noise vectors to generate intermediate flows.

\paragraph{Computational cost} 
The use of emotion-specific flow steps reduces computational overhead during inference, thereby improving the overall generation speed. On an NVIDIA RTX A5000, we measured the inference time required to generate a single audio sample of fixed length, using the same reference audio and text input across settings. 
As shown in Table~\ref{tab:ctrlnet_inference_time}, applying CtrlNet only to the early flow steps $[0,\ 0.1]$ introduces a small overhead compared to the baseline, while applying it to the full range $[0,\ 1]$ increases inference time more noticeably. This demonstrates that partial control offers a good trade-off between performance and efficiency.

These values were averaged over 20 runs to ensure consistency. The results demonstrate that partial application of CtrlNet in early flow steps is more efficient than full-step control, effectively balancing inference speed and emotional expressiveness.

\begin{table}[h]
\centering
\caption{Inference time (seconds) with different CtrlNet application ranges.}
\resizebox{0.5\linewidth}{!}{\begin{tabular}{l c}
\toprule
\textbf{CtrlNet application range} & \textbf{Inference time (s)} \\
\midrule
No CtrlNet (baseline)             & 3.708 \\
CtrlNet on $[0,\ 0.1]$            & 4.212 \\
CtrlNet on $[0,\ 1]$              & 5.416 \\
\bottomrule
\end{tabular}}
\label{tab:ctrlnet_inference_time}
\end{table}

\newpage
\section{More explanation of emotion window size}
\label{asec:emotion_window_size}
\begin{figure}[t]
  \centering
\includegraphics[width=0.85\linewidth]{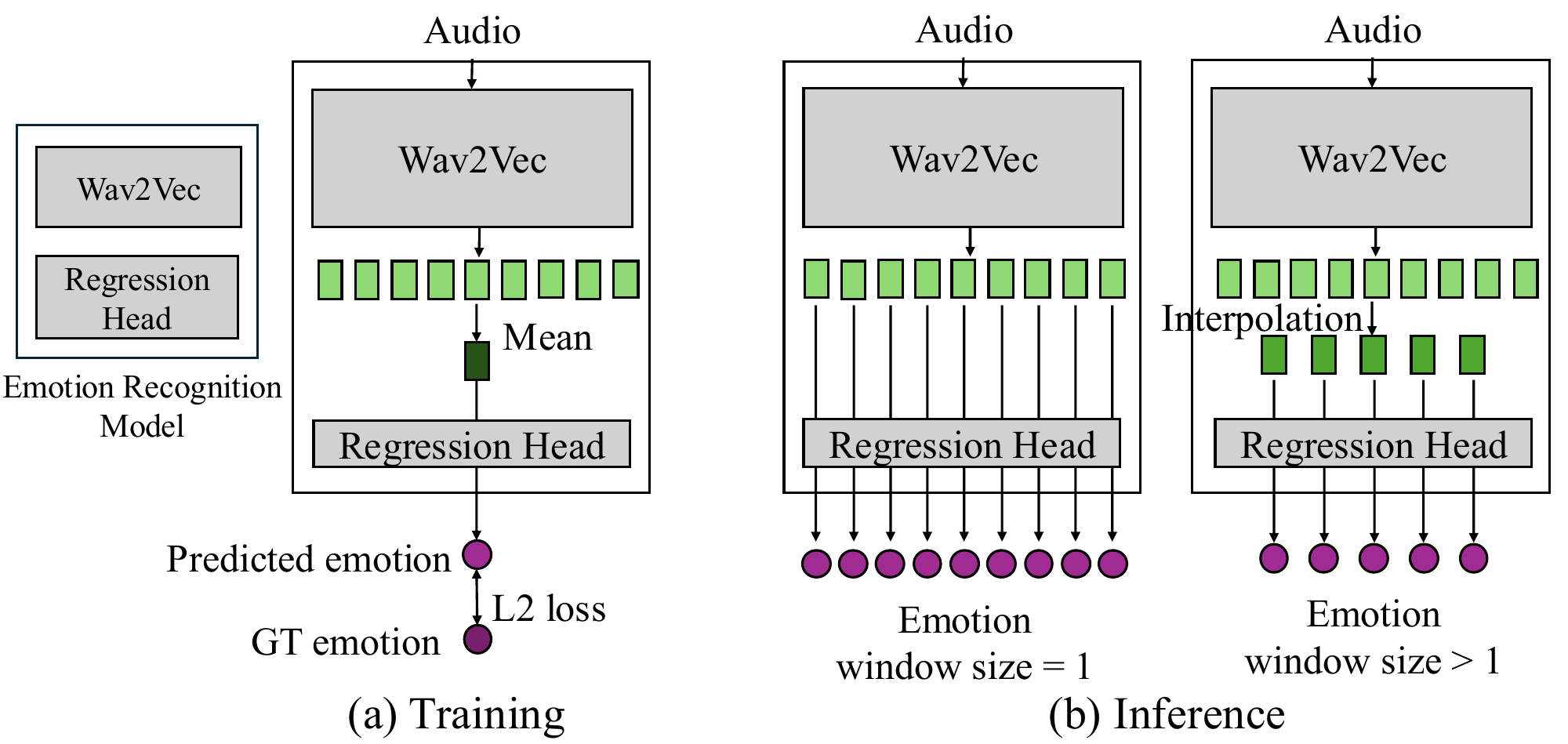}
  \caption{\textbf{Emotion window step} Using window size = 1 fails to convey emotion properly, while setting a proper window size improves prediction consistency}
  \label{fig:expresion_interpolation}
\end{figure}

In \sref{sec:emotino_window_size}, we propose not to use emotion window size = 1 in the speech emotion recognition (SER) model \cite{wagner2023dawn}. For better clarity, we also include a corresponding \fref{fig:expresion_interpolation} to illustrate this setup. During training, the SER model is trained on full utterances paired with a single emotion label, since it is challenging to annotate time-varying emotions. For a given audio input, the model uses wav2vec to produce a sequence of token-level features, averages them, and then passes the result through a regression head to predict a single emotion.
At inference time, we mirror this training setup by applying interpolation over wav2vec outputs using an emotion window size greater than 1, effectively approximating the averaging process. This leads to improved word error rate (WER) and emotion similarity compared to using no interpolation (i.e., window size = 1), as shown in Table 2.

\newpage
\newpage

\medskip


\newpage

\end{document}